\documentclass[aps,prl,twocolumn,epsfig,floats]{revtex4}
\usepackage{graphicx,epsf,natbib}
\def\be{\begin{equation}}
\def\ee{\end{equation}}
\def\bea{\begin{eqnarray}}
\def\eea{\end{eqnarray}}

\begin{document}


\title{Theory of the Quantum Critical Fluctuations in Cuprates}

\author{Vivek Aji and C. M. Varma}
\address{Physics Department, University of California,
Riverside, CA 92507}
\begin{abstract}
 The statistical mechanics of the time-reversal and inversion symmetry breaking order parameter, possibly observed in the pseudogap region of the phase diagram of the Cuprates,
 can be represented by the Ashkin-Teller model. We add kinetic energy and dissipation to the model for a quantum generalization and show that the correlations are determined by
 two sets of charges, one interacting locally in time and logarithmically in space and the other locally in space and logarithmically in time. The quantum critical fluctuations
 are derived and shown to be of the form postulated in 1989 to give the marginal fermi-liquid properties. The model solved and the methods devised are likely to be of interest
 also to other quantum phase transitions.

\end{abstract}
\maketitle

  A mean-field solution of a microscopic model predicts that the {\it pseudogapped} region \cite{norman} of the phase diagram of the
  cuprates (Region II in fig.(1) of Ref. (2)) has a spontaneous current pattern depicted in fig.(1a) \cite{cmv}. Such a state breaks time-reversal,
  inversion and three of the four reflection symmetries of the square lattice while preserving translational
 symmetry.  The observed magnetic diffraction in neutron scattering\cite{FAQ} as well as the dichroism observed in
 ARPES \cite{AK} with circularly polarized photons is consistent with such a state.

  The transition temperature of  this state varies continuously with hole density $x$ and tends to $0$ at $x=x_c$, the quantum critical point. Anomalous normal state properties
  are observed in the funnel shaped region (Region I in fig.(1) of Ref. (2)) emanating from the QCP. They were shown to
 follow from a  phenomenological spectrum \cite{mfl}:
 \begin{eqnarray}\label{mfl}
     \text{Im}\chi({\bf q}, \omega, T) & \propto & \omega/T, ~ for~ \omega \ll T,   \\  \nonumber
                                              & \propto & constant, ~ for ~ T \ll \omega \ll \omega_c,
    \end{eqnarray}
  where  $\omega_c$ is a cut-off, determinable from experiments. The corresponding real part of the fluctuations, $Re\chi({\bf q}, \omega, T) \propto \ln(\omega_c/\omega)$
  for $\omega/T \gg 1$ and $\propto \ln(\omega_c/T)$ for $\omega/T \ll 1$.  Since this spectrum has a singularity in the limit $(T,\omega) \to 0$, it specifies the fluctuations
  near a quantum critical  point. For small ${\bf q}$, this spectrum is directly observed in Raman scattering. \cite{raman}. The two unusual properties of Eq.\ref{mfl}, $\omega/T$
  scaling \cite{sachdev} and no dependence (or a smooth dependence) on ${\bf q}$ are sufficient to generate a marginal fermi-liquid \cite{mfl}. The observed anomalous properties are well explained by the marginal fermi-liquid and its predictions for the single-particle spectra have been verifed in experiments.

The purpose of this work is to show that the quantum critical spectrum
 of Eq.(\ref{mfl}) is the spectrum of fluctuations of the order parameter which condense to give the observed order depicted in fig.(1a).

\begin{figure}[ht]
  \centering
  \includegraphics[width=0.8\columnwidth]{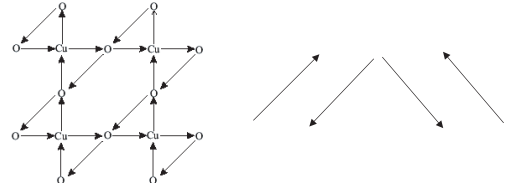}
 \caption{ Current pattern(a) and its abstraction (b)}
 \label{ccp}
\end{figure}
To do so we consider a quantum generalization of
 the  classical model
 whose solution gives such an order parameter. The four possible configurations in each unit-cell of fig.(1a) can be represented by four vectors with arrows representing
 time-reversal and orientation representing
 the only plane of reflection symmetry, as shown in fig.(1b). A classical model with such a pair of Ising degrees of freedom is the Ashkin-Teller \cite{BAX} model:
 \begin{equation}\label{d-ising}
 H = -\sum_{<i,j>}J_2(\sigma_i \sigma_j + \tau_i \tau_j) +J_4(\sigma_i  \tau_i \sigma_j \tau_j)
  \end{equation}
  $\sigma_i$ and $\tau_i$ are Ising spins.
  Such a Hamiltonian can also be derived \cite{cmv}  from the microscopic model which in mean-field theory gives the observed order parameter provided one neglects amplitude
  fluctuations.

  The Ashkin-Teller model has a variety of phases depending on $J_4/J_2$ and $T/J_2$ \cite{BAX, KD, MML}.
  Especially interesting to us is the region $-1<J_4/J_2<0$ in which the ordered phase has ferromagnetic order in $<\sigma_i>, <\tau_i>$ as well as $<\sigma_i\tau_i>$ just as in
  the order observed. Two more properties of this region are of special interest: on the critical line, the model is Gaussian (with stiffness depending on $J_4/J_2$) and the
  specific heat at the transition has no divergence.

  With the transformation
  \be
  \label{softAT}
\cos(\theta_i) = (\sigma_i+\tau_i)/2;~\sin(\theta_i) =
(\sigma_i-\tau_i)/2, \ee
 the Ashkin-Teller model becomes
\be\label{ast} H = \sum_{i,j nn} 2 J_2 \cos(\theta_i-\theta_j) + J_4
\cos 2(\theta_i-\theta_j) + h_{4} \cos 4\theta_i.
\ee
The
restriction that  $\theta_i = 0,\pi/2,\pi,3\pi/2$ has been written
in terms of a four-fold anisotropy term $\propto h_{4}$. Monte-Carlo calculations show \cite{sudbo} that the classical phase diagram of model (\ref{d-ising}) and of
(\ref{softAT}) are the same. This is consistent with the fact that $h_{4}$ is irrelevant above the critical point in the $xy$ model; $h_4$ however is relevant below the
transition enforcing an Ising state \cite{JOS}.

For a quantum theory for the model, two additional features are included, the kinetic energy and the
dissipation. For the latter, we consider physics of the
Caldeira-Leggett type \cite{CL}, which can be derived by integrating
over the fermions in the microscopic model which couples the
orbital-currents of $(\theta_i-\theta_j)$ to the currents of the
fermions \cite{footnoteDISS}. For simplicity of presentation, we start with $J_4=0$. We
will show that no essential modification arises at least for
$|J_4|/J_2 <1/2$, which is in the interesting regime. We will also
take $h_4=0$ and show that in the fluctuation regime, which is our
primary interest, a finite $h_4$ does not change the results. With
these simplifications,

\begin{eqnarray}\label{mod}
S &=& S_{Q} + \int \sum_{<ij>} d\tau 2J_2
\cos\left(\theta_{i}\left(\tau\right)-\theta_{j}\left(\tau\right)\right),
\\ \nonumber
S_{Q} &=& \int d\tau\sum_{i}{1\over E}\left(d\theta_{i}\over
{d\tau}\right)^{2} + \\ \nonumber  & &\int d\tau \int d\tau
'\sum_{<ij>} {\alpha\over
{8\pi}}\left({\left(\theta_{ij}\left(\tau\right)-\theta_{ij}\left(\tau
'\right)\right)\over {\left(\tau - \tau '\right)}}\right)^{2}
\end{eqnarray}

\noindent where $\theta_{ij}\left(\tau\right) =
\theta_{i}\left(\tau\right)-\theta_{j}\left(\tau\right)$.

We use the Villain representation of the periodic functions and introduce
integer link variables $\textbf{m}_{i} = m_{i x}\widehat{x}+m_{i
y}\widehat{y}$ and discretize time in steps $\Delta\tau$ to get
\begin{eqnarray}
S &=& \sum_{\left<\mu\nu\right>}\sum_{i}{1\over
E\Delta\tau}\left(\theta_{i\mu}-\theta_{i\nu}\right)^{2}+ S_{diss}
\\ \nonumber &+& \sum_{\left<ij\right>}\sum_{\mu}2J_2\Delta\tau\left(\theta_{i\mu}-\theta_{j\mu}-2
\pi m_{ij}\right)^{2}
\end{eqnarray}

After a Fourier transform and  integration over the $\theta_{ln}$
variables, the action for the vector field $\textbf{m}$ is obtained:

\begin{eqnarray}
S &=& {1\over {L^{2}\beta}}\sum_{ln}\left(2\pi\right)^{2}
G\left(\textbf{k}_{l},\omega_{n}\right)\left[ {J a^{4}
\left|\textbf{k}_{l}\times \textbf{m}_{ln}\right|^{2}}\right.
\nonumber \\ &+& \left. {\left(J_{t}+ {\alpha\over {4\pi
c}}{k_{l}^{2}\over {\left|\omega_{n}\right|}} \right)a^{2}\left(\Delta\tau\right)^{2}|\omega_{n}{\bf m}_{ln}|^{2}
}\right]
\end{eqnarray}

\noindent  Here $\textbf{k}_{l}$ are discretized momenta, $\textbf{k}_{l}= (2\pi l_{x}/a)\hat{x}$ $ +
(2\pi l_{y}/a)\hat{y}$, $\omega_{n}$ are the Matsubara frequencies, $\omega_{n} = 2\pi n/\beta$, and $J =
2J_{2}\Delta\tau/a $ , $ J_{t} = 1/Ea\Delta\tau $. The propagator $G\left(\textbf{k}_{l},\omega_{n}\right)$ is given by
\be
G\left(\textbf{k}_{l},\omega_{n}\right) = \left(J c k_{l}^{2}
 + J_{t}\omega_{n}^{2}/c +
\alpha\left|\omega_{n}\right|k_{l}^{2}/4\pi\right)^{-1}.
\ee

Quantum dynamics introduces sources and sinks in the vector field.
Therefore this action includes besides the vortices, (curl
$\textbf{m}$), the time-derivatives of $\textbf{m}$. Sources and
sinks due to quantum dissipation suggest that the time-derivative
must include a field with a
  divergence. We define two
  {\it orthogonal} charges,
$\rho_{v}(\bf{r},t)$
and $\rho_{0}(\bf{r},t)$ (in the continuum limit) through
\be
\label{charge1}
\rho_{v}(\bf{r},t)\hat{\bf{z}} = \nabla \times
\bf{m}(\bf{r},t),
\ee
\be
\label{charge2}
 \rho_{0}({\bf r},t)=(1/c)d{\bf m}({\bf r},t)/dt \cdot
\hat{\bf{r}}
\ee

\noindent where the velocity $c = a/\Delta\tau$. The time derivative
of $\textbf{m}$ appearing in the action then is
\begin{equation}
\omega_{n}^{2}\left|\textbf{m}\right|^{2} = {\omega_{n}^{2}\over
k_{l}^{2}}\left|\rho_{v}\right|^{2} + c^{2}\left|\rho_{0}\right|^{2}.
\end{equation}

\noindent In the theory of the $xy$ model,
 the charges $\rho_v$ are sources of "vortices" where the velocity field is
azimuthal with a strength falling off as $1/r$. $\rho_{0}$ are sources of the orthogonal radial
field \cite{footnote1}. From (\ref{charge2}), it is seen that $\rho_0$'s are events in time where the
divergence of the field ${\bf m}({\bf r}, t)$ changes. The importance of $\rho_{0}$
lie in their logarithmic interaction in time.

In terms of  $\rho_v, \rho_0$, the action, in the continuum limit, neatly
splits into three parts: $S=S_{v} + S_{0} + S'_{0}$

\begin{eqnarray}\label{chgact}
S_{v} &=& {1\over {L^{2}\beta}}\sum
{J\over {c k_{l}^{2}}}\left|\rho_{v} (\textbf{k}_{l}\omega_{n})\right|^{2}\\
\nonumber S_{0} &=& {1\over {L^{2}\beta}}\sum{\alpha\over
{4\pi\left|\omega\right|}}\left|\rho_{0}(\textbf{k}_{l}\omega_{n})\right|^{2}
\\ \nonumber S'_{0} &=& {1\over {L^{2}\beta}}\sum
G\left(\textbf{k}_{l},\omega_{n}\right) \left(J J_{t} - {\alpha
J_{t}\left|\omega_{n}\right|\over {4\pi c}} -
{\alpha^{2}k_{l}^{2}\over {16\pi^{2}}}\right)\\ \nonumber && \times
\left|\rho_{0}(\textbf{k}_{l}\omega_{n})\right|^{2}
\end{eqnarray}

The interesting part about this decomposition \cite{footnote(nondissipative)} is that
$S_v$ is the Coulomb gas representation of the $xy$ model with charges interacting logarithmically in space
but locally in time and  $S_0$ is the Coulomb gas representation of the Kondo problem
with charges interacting logarithmically in time but locally in space.  $S'_{0}$
has non-singular interactions and is unimportant compared to $S_v$
and $S_0$ \cite{footnotex}. Being orthogonal the charges $\rho_v$ and $\rho_0$ are
uncoupled; the action is a product of the action over configurations
of $\rho_v$ and of $\rho_0$. Any physical correlations are
determined by correlations of both charges. Both the $\left<\rho_v\rho_v\right>({\bf k},\omega)$ and the $\left<\rho_0\rho_0\right>({\bf k},\omega)$
correlations are well understood. For any finite $J$, the field $\rho_{v}$ is
confined in the limit $T\to 0$; no free vortices exist. The phase
transition can come about only due to free $\rho_0$'s, due to a
tuning of dissipation parameter $\alpha$. We therefore first remind ourselves of the correlation function of
the $\rho_0$'s.

Let us introduce a core-energy $\Delta$ for the
$\rho_0$'s just as is done to control the fugacity of $\rho_v$'s, the vortices.
Next consider
 how the renormalization of $\alpha$ and
$\Delta$ proceeds.
\noindent Including the core-energy, the action $S_0$ is
\begin{equation}
S_0 = \sum_{i}\left[T\sum_{n}{\alpha\over {4\pi}}{1\over
\left|\omega_{n}\right|}\left|\rho_{0i}\left(\omega_{n}\right)\right|^{2}
+ \int d \tau
\Delta\left|\rho_{0i}\left(\tau\right)\right|^{2}\right]
\end{equation}

 This is identical to the Coulomb gas representation \cite{AHY}
of the Kondo problem and a quantum dissipation problem \cite{LEG}. The
resulting RG equations are
\begin{eqnarray}\label{rgflow}
{d \widetilde{\Delta}\over {dl}} &=& \left(1 - \alpha\right)\widetilde{\Delta}\\
\nonumber {d\alpha \over {dl}} &=& -\alpha \widetilde{\Delta}^{2}
\end{eqnarray}
\noindent where $\widetilde{\Delta} = \Delta\tau_{c}$ and
$\tau_{c}$ is a short time cutoff. \noindent The critical point of
interest is at $\alpha_{c} = 1$ \cite{footnote2},  where $\widetilde{\Delta}$ scales to 0; for
$\alpha <1$ the charges $\rho_0$ freely proliferate as "screening"
due to $\widetilde{\Delta}$ becomes effective. $\alpha >1$
represents the ordered or confined region in which the anisotropy
field $h_4$ is strongly relevant. We are interested here only in the
region $\alpha \leq 1$. Well in the quantum critical region, the
(singular part of the) propagator for $\rho_0$ is \be
\label{rho0corr}
\left<\rho_{0}\left(\omega_{n}\right)\rho_{0}\left(-\omega_{n}\right)\right>
= \left(1/4\pi\left|\omega_{n }\right|\tau_{c}\right)^{-1}. \ee The
crossover to the quantum-disordered or screened state is given when
$\omega_x$ is of the order of the inverse of the characteristic
screening time, which may be estimated similarly to Kosterlitz's
estimate \cite{Kosterlitz} of the screening length in the
$xy$-problem. Thus
 \be \label{crossover} \omega_x \approx
\tau_c^{-1} \exp\left(- b/\sqrt{1-\alpha}\right), \ee
where b is a
numerical constant of O(1). At finite temperatures and low
frequencies, the crossover temperature may be taken as given by
Eq.(\ref{crossover} ) with $\omega_x$ replaced by $T_x$.

Finally, we come to the correlation function of interest, that of the order parameter:
\begin{equation}
\label{orderparameter}
C_{i,j,\mu,\nu} = \left<e^{\imath \theta_{i\mu}}e^{-\imath\theta_{j\nu}}\right> \equiv {
\int d\left[\theta\right]e^{-\bar{S}}\over {\int
d\left[\theta\right]e^{-S}}}
\end{equation}
\noindent To compute this we will employ a procedure similar to the
one used by Jose et al. \cite{JOS} for the 2d $xy$ model. Consider a path from
$i\mu$ to $j\nu$, and split it into two: $i\mu \rightarrow j\mu$ and
$j\mu \rightarrow j\nu$. All paths should give the same answer so we have
chosen the one most convenient. A
 vector field $\overrightarrow{\eta}_{i\mu}$ is defined which lives on
the sites of the lattice and whose components are $1$ if the path
crosses site $i\mu$ and zero otherwise. To capture
the second part of the path, we define a scalar field
$\eta^{0}_{i\mu}$, which lives on the links in time, and is nonzero
only for paths on the link.

\noindent Including the fields described above and integrating over the $\theta's$, we get

\bea
\bar{S} = S &+& {1\over {L^{2}\beta}}\sum_{ln}\left(2\pi\right)J
G\left(\textbf{k}_{l},\omega_{n}\right)\Large[\left(\omega_{n}\eta^{0\star}/2\right)\left(\imath\bf{k}_{l}\cdot\bf{m}_{ln}\right)
 \nonumber \\  &\ & +
\left(\textbf{k}_{l}\cdot\overrightarrow{\eta}^{\star}/2\right)\left(\imath\textbf{k}_{l}\cdot\textbf{m}_{ln}\right) +c.c. \nonumber \\ &\ & +
 \left|\omega_{n}\eta^{0}/2+\textbf{k}_{l}\cdot\overrightarrow{\eta}^{\star}/2\right|^{2}\Large].
\eea

\noindent The last term in the sum gives the spin wave contribution to the
correlation function. Clearly the spin wave contribution is finite
and this is indeed the standard result that spin wave fluctuations
do not disorder the state in three dimensions. The second term in
the sum can be written in terms of the linear coupling of
(${\bf k}\times \overrightarrow{\eta}$) to $\rho_v$'s, the vortices. The vortices are confined for $T\to 0$ for the case of a finite $J$ assumed by us and
are therefore not interesting. The first term can be written in terms of a linear coupling between $\eta^0$'s and $\rho_0$'s. This is the interesting term because the
  divergence in frequency induced by the
dissipation as $\alpha$ is tuned leads to a proliferation of  $\rho_0$'s and thus to disorder.
The correlation function  near the quantum critical point are thus
determined entirely by the  dynamics of $\rho_0$'s. The
correlation function is proportional to,

\begin{eqnarray}\label{gv}
&&C({\bf r-r'}, \tau-\tau') = \exp (-F), \\
\nonumber &&F = -{ J^{2}c^{2}\over 4}T\sum_{n}\int
{d\textbf{k}}{k^{2}\over
{\omega_{n}^{2}}}\left<\rho_{0}\left(\omega_{n}\right)\rho_{0}\left(-\omega_{n}\right)\right>\times  \\
\nonumber && G^{2}\left(\textbf{k},\omega_{n}\right)\left(1-
\cos\left(\textbf{k}\cdot\left(\textbf{r}-\textbf{r}'\right)+\omega_{n}\left(\tau-\tau'\right)\right)
\right).
\end{eqnarray}

Deep in the quantum critical regime the correlations function
$\left<\rho_{0}\left(\omega_{n}\right)\rho_{0}\left(-\omega_{n}\right)\right>$
is given by Eq.(\ref{rho0corr}). The summand over $n$ in $F$ has a
leading $1/\left|\omega\right|$ part. It is easy to see that for any
finite $\left|\textbf{r}-\textbf{r}'\right|$ the sum over $n$ is
divergent. Thus, for any finite spatial separation, the correlation function
$C({\bf r-r'}, \tau)$
is identically zero for $\tau$ larger than the
crossover scale $\tau_c$. On the other hand, for
$\left|\textbf{r}-\textbf{r}'\right| = 0$, the correlation
 function is given
by $C(0,\tau-\tau')=\exp(-F(0,\tau-\tau'))$, where
\begin{equation}\label{gv}
 F(0,\tau-\tau')=-{ 2\pi T}\sum_{n} {1-
\cos\left(\omega_{n}\left(\tau-\tau'\right)\right)\over
\left|\omega_{n}\right|}\log\left(
{\left|\omega_{n}\right|\tau_{c}}\right).
\end{equation}

Such correlation functions have been calculated in other contexts.
In particular Ghaemi et al. \cite{GAS}, provide the spectral
function of the correlation to be

\be \label{spectra} Im \chi({\bf
q},\omega) = \overline{\tau_c}\tanh(\omega/2T). \ee

 \noindent where
$\overline{\tau_{c}} = c_{0}\tau_{c}$, $c_{0}$ is a constant of
$O(1)$. The phenomenological spectrum of Eq.(\ref{mfl}) is thus derived.

We now summarize the calculations for the effect of the anisotropy field $h_{4}$.
As in Ref.(\cite{JOS}), we  supplement the Action of Eq. (\ref{chgact}) by  introducing a new field $p_{i}$
\begin{equation}
e^{h_{4}\cos\left(4\theta_{i}\right)}\approx
\sum_{p_{i}}e^{\ln\left(y_{4}\right)p_{i}^{2} + \imath 4
p_{i}\theta_{i}}
\end{equation}
\noindent where $y_{4} = h_{4}/2 $. Introducing the $\textbf{m}$
fields as before and performing the $\theta$ integral leads to
additional terms in the action which couple $p_{i}$ linearly to $\rho_{0,i}$. This linear coupling can be eliminated with a renormalization of the coupling
constant $\alpha$ to $2\alpha$. This does not change the critical behavior in the quantum critical regime; the correlation remains of the form (\ref{spectra}).
For $\alpha > \alpha_c$, $h_4$ is relevant just as in the classical problem and enforces an Ising state.

The coupling $J_{4}$ is easier to treat. For  $-0.5<J_4/J_2<0$, the bare potential  continues to have the absolute  minima at $(\theta_i-\theta_j)=0$.
Therefore the Villain representation of periodic functions can be made as above with an altered coefficient. This affects the cut-offs in the solution
but not the form of the correlation function below the cut-offs. The theory is valid only in this range; for smaller $J_4$, an Ising transition to a different
symmetry is expected as in the classical model.

We summarize the principal results: We have considered a statistical mechanical model which has the symmetries and the degeneracies of the observed phase in
underdoped cuprates. The specific heat at the classical transition in this model is a smooth function of temperature as in experiments. We have generalized
the model by including inertial as well as dissipative dynamics. On tuning dissipation, the model has a phase transition at $T=0$. The critical fluctuations
of the model are determined by the dynamics of charges with interactions which are spatially local but logarithmic in time. The correlation function in the
quantum critical regime calculated for the model has the form postulated phenomenologically to understand the properties of cuprates in the quantum critical
regime, i.e they display $\omega/T$ scaling and spatial locality.

More generally, our results are applicable also to the dissipative quantum $xy$ model and hence relevant to critical properties of Josephson Junction arrays
in two dimensions. The Ashkin-Teller model is a staggered 8-vertex model. We expect that our method has
application to quantum versions of 6 and 8-vertex models generally to which many physical problems of interest correspond.

We wish to thank T. Giamarchi, B.I. Halperin, A. Shehter, A. Sudbo,
and A. Vishwanath for useful discussions.


\begin{thebibliography}{10}
\bibitem{footnoteDISS}
To check the consistency of integration over fermions, one must make sure that the dissipative coupling has no essential change when it is calculated with the renormalized fermions obtained by coupling to the fluctuations calculated here.
\bibitem{footnote1}
Note that $\rho_{0}$ is not proportional to the divergence of ${\bf m}$. On the other hand,
only the part of ${\bf m}$ with a divergence contributes to $\rho_{0}$. The velocity field from
$\rho_{0}$ at the origin falls off as $1/r^{2}$.
\bibitem{footnotex}
This has also been explicitly tested by calculating the correlations,
 Eq.(\ref{orderparameter}), with and without $S'_{0}$.
\end{thebibliography}
\end{document}